\documentclass[twocolumn,preprintnumbers,amsmath,amssymb]{revtex4}
\usepackage{graphicx}
\usepackage{dcolumn}
\usepackage{bm}
\usepackage[usenames,dvipsnames]{xcolor}

\begin{document}

\title{Magnetic ghosts and monopoles}

\author{N.Vandewalle}
\affiliation{GRASP, Physics Department, University of Li\`ege, B-4000 Li\`ege, Belgium.}
\author{S.Dorbolo}
\affiliation{GRASP, Physics Department, University of Li\`ege, B-4000 Li\`ege, Belgium.}

\date{\today}
 
\begin{abstract}
While the physics of equilibrium systems composed of many particles is well known, the interplay between small-scale physics and global properties is still a mystery for athermal systems. Non-trivial patterns and metastable states are often reached in those systems. We explored the various arrangements adopted by magnetic beads along chains and rings. Here, we show that it is possible to create mechanically stable defects in dipole arrangements keeping the memory of dipole frustration. Such defects, nicknamed ``ghost junctions", seem to act as macroscopic magnetic monopoles, in a way reminiscent of spin ice systems.
\end{abstract}


\maketitle

\begin{figure}[h]
\begin{center}
\includegraphics[width=8.5cm]{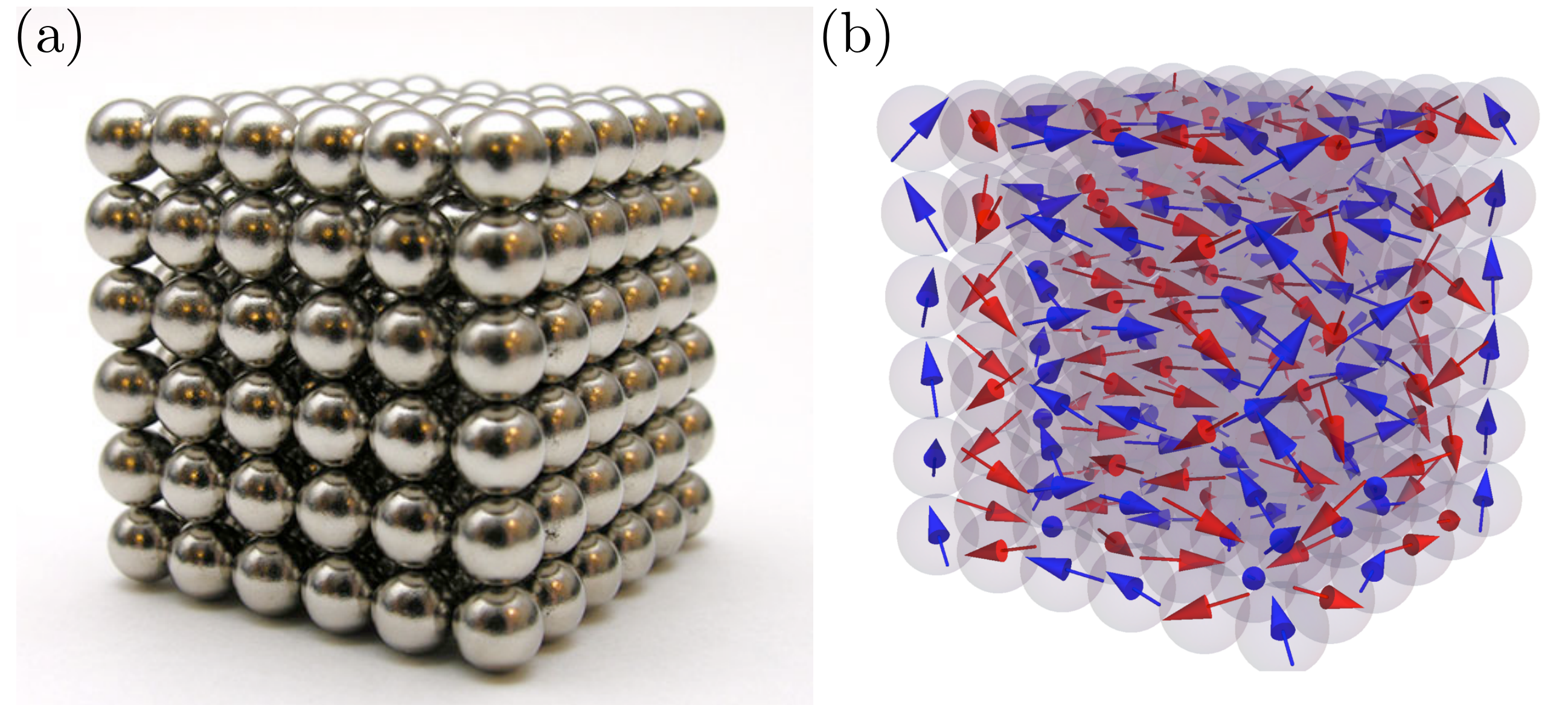}
\vskip -0.2 cm
\caption{(a) A popular puzzle : a cube of $6^3$ magnetized beads. (b) Dipole ordering in the cube as obtained from our numerical simulations based on the minimization of total energy $U$. Blue and red colors are used to distinguish upward/downward orientations of the dipoles along the vertical axis. }
\label{fig_cube}
\end{center}
\end{figure}

Neodyme sphere magnets are a beloved puzzle for geeks. Since dipole-dipole interactions are stronger than the weight of the beads, stable structures such as chains (1D), hexagons (2D) and cubic lattices (3D) can be easily created. Figure \ref{fig_cube}(a) presents a cube composed by 216 beads. Following tips and tricks, complex 3D structures can be also built from icosahedra to fractal Sierpinsky pyramids. Behind the broad variety of amazing structures that it is possible to build with magnetized beads, dipole-dipole interactions are more and more proposed to generate self-assembled structures at the mesoscopic scale \cite{selfassembly,pelesko,pnas,aranson,softmatter}, using for example magnetic colloids \cite{fermigier,maret}. The study of macroscopic dipoles is therefore relevant for many applications at different scales. 

While the dipole ordering was deeply studied in the case of equilibrium systems \cite{kadanoff}, the case of athermal dipoles was poorly explored. Only a few experiments were performed in order to study the various configurations adopted by a collection of magnetized beads. In a pioneering study, Blair and Kudrolli \cite{blair} realized series of experiments onto a vibrating plate, injecting mechanical energy into the system : chains, rings, and 2D crystals have been observed. Lumay and Vandewalle \cite{tunable} explored the properties of a granular packing submitted to a vertical magnetic field : the beads are organised such that low packing fractions can be reached. In another experiment, Carvente and Ruiz-Suarez \cite{carvente} obtained denser self-assembled systems using magnetized spheres. In the dilute limit, Falcon and coworkers \cite{falcon} experimented random magnetic forcing of a granular gas.

The main motivation of the present work is to explore the possible dipole configurations adopted by a collection of magnetized spheres. A series of fundamental questions arises : What is the link between the stability of an assembly composed of several magnetic beads and the dipole orientations ?  Do different (metastable) states exist ? We performed several experiments with magnetized beads and we rationalized results using numerical simulations. In this paper, we present the striking results obtained with apparent simple systems. The most relevant one is the obtention of monopole-like behaviours.

As found in classical textbooks \cite{jackson}, uniformly magnetized spheres behave like dipoles. The interaction energy between two point-like dipoles $\vec m_i$ and $\vec m_j$ is given by
\begin{equation}
u_{ij}= {\mu_0 \over 4\pi} \left[ {\vec m_i . \vec m_j \over r_{ij}^3} - 3 {(\vec m_i . \vec r_{ij})(\vec m_j . \vec r_{ij}) \over r_{ij}^5}\right] .
\label{eq_dipole}
\end{equation}
where $\vec r_{ij} = \vec r_j - \vec r_i$ is the vector linking particles $i$ and $j$. We consider identical beads such that they have similar sizes and similar magnetizations ($| \vec m_i | = m$). It is therefore possible to define a dimensionless macroscopic potential as
\begin{equation}
U={2\pi D^3 \over \mu_0 m^2} \sum_{i \ne j} u_{ij} ,
\end{equation}
where $D$ is the sphere diameter and $m$ being the bead moment. The algorithm used in our work consider the positions $(x_i,y_i,z_i)$ and angular orientations $(\theta_i,\varphi_i)$ of each dipole $i$. In order to explore different structures, the sphere positions are fixed while the orientation of dipoles are free parameters. The algorithm starts from a random orientations of the spins and searches iteratively for the minimum of energy by changing slightly the angles $\theta_i$ and $\varphi_i$.  As a first example, Figure \ref{fig_cube}(b) proposes one of the low energy configurations for the spins arranged in a cube. Colors indicate different dipole orientations along the vertical axis. A complex ordering is found inside the cube. Along the main axes of the cube, chains of dipoles having similar orientations are found. Moreover, helicoidal-like orientations are also observed.  

\begin{figure}[h]
\begin{center}
\includegraphics[width=6.0cm]{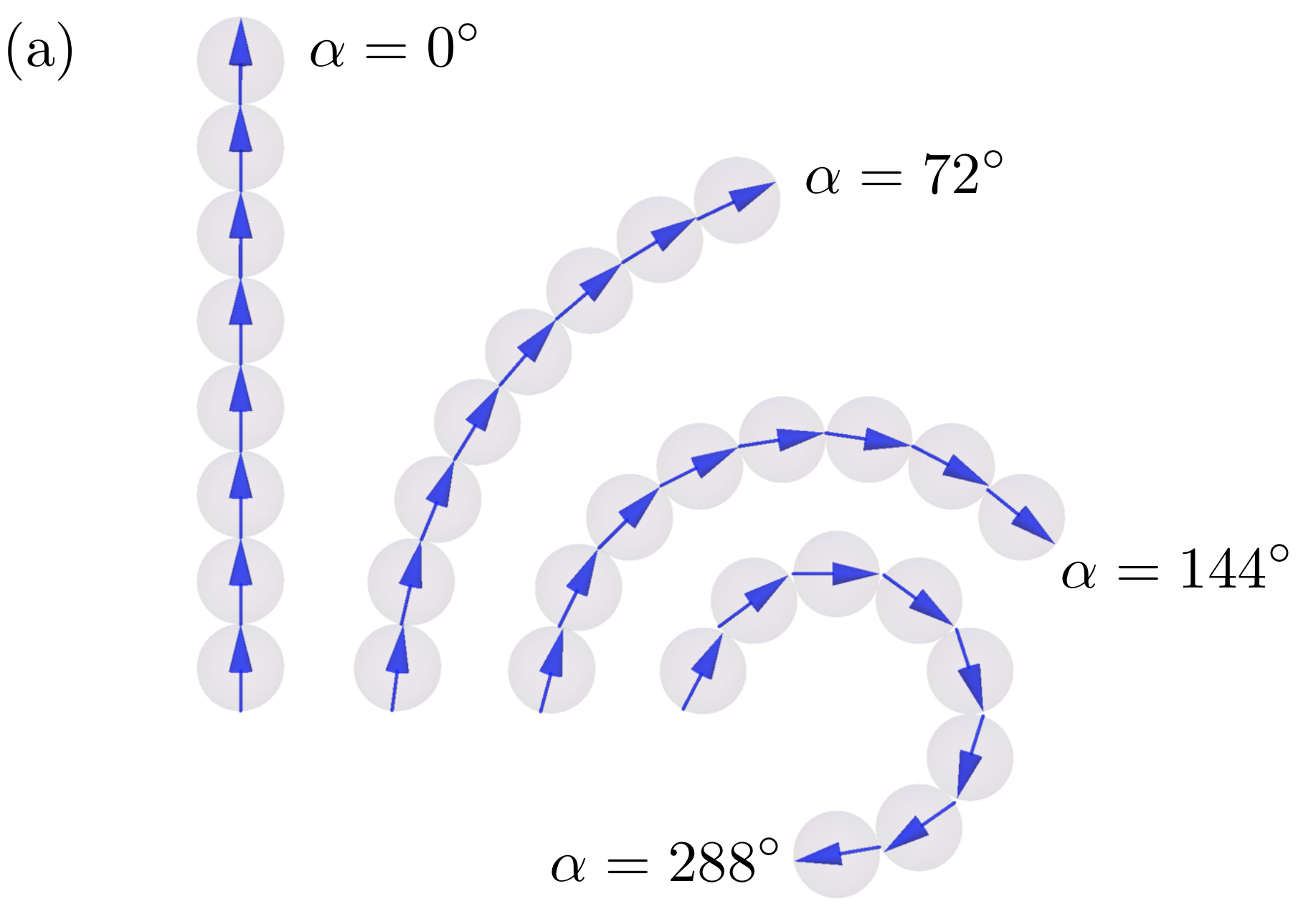}
\includegraphics[width=7.0cm]{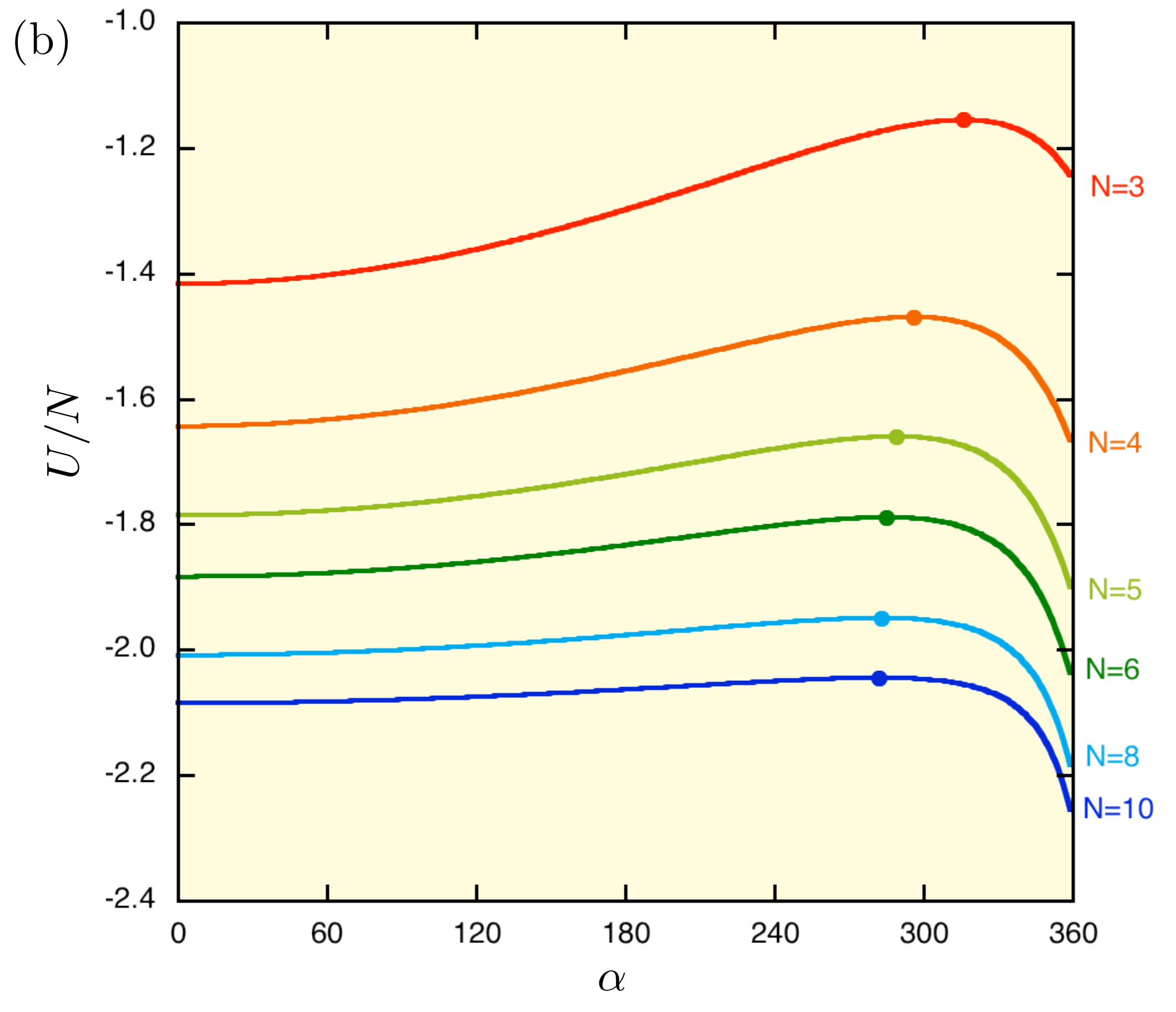}
\includegraphics[width=8.5cm]{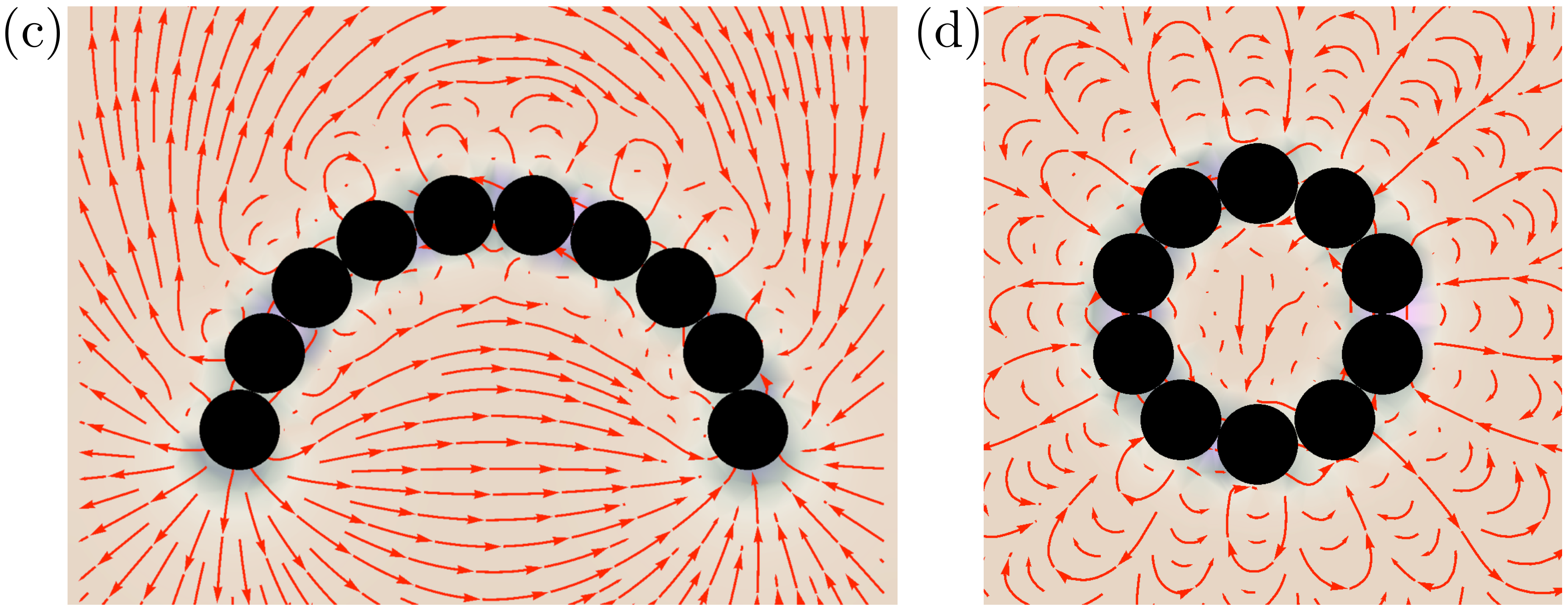}
\vskip -0.2 cm
\caption{(a) A chain of $N=8$ dipoles bent from $\alpha=0$ (line) to $\alpha=360^\circ$ (closed ring). (b) Energy per particule $U/N$ for a chain continuously bent from 0 to an angle $\alpha$. Different chain sizes $N$ are illustrated. On each curve, the dot indicates the maximum value giving the barrier position. (c) Field lines around a curved chain of $N=10$ beads as obtained in numerical simulations :  a dipole-like structure is seen, the extremities of the chain being the source and sink of field lines. (d) Field lines around a ring of 10 beads. A multipole structure is observed, each sphere being the source and the sink of field lines.  }
\label{fig_chain}
\end{center}
\end{figure}

Since 3D structures show complex dipole ordering, we first focused on chains of magnetic beads, as shown in Figure \ref{fig_chain}(a). Chains are known to represent the natural way magnetic particles self-assemble. Indeed, the interaction given by Eq.(\ref{eq_dipole}) is highly anisotropic, and strong attractive interactions are obtained for aligned dipoles \cite{rosenweig}. By searching the minimization of energy of the dipoles, the ground state is found to be a simple alignment of the dipoles along the chain, as shown in Figure \ref{fig_chain}(a). One has
\begin{equation}
U_0 = -2 \sum_{i=1}^{N-1} {(N-i) \over i^3}
\label{eq_u0}
\end{equation}
From that pattern, the continuous deformation of a chain towards a ring configuration will emphasize the transition seen in Ref. \cite{blair}. The chain is slightly deformed such that it forms an arc with a cumulated angle $\alpha$. When $\alpha$ reaches $360^\circ$, a ring is formed. The radius of curvature of the chain is therefore $R= N D / \alpha$. From numerical simulations, Figure  \ref{fig_chain}(b) shows the energy per dipole $U/N$ as a function of $\alpha$ for different chain sizes $N$.  Two minima are seen in the curves at $\alpha=0$ (chain) and $\alpha=360^\circ$ (ring) respectively. They are separated by an energy barrier, whose maximum is around $\alpha_{max} \approx 280^\circ$ for large $N$ values. This energy barrier can be tested experimentally as shown in the supplementary movie. For finite systems (and for $N>3$), the ring configuration is more stable than the chain. This explains why rings were often observed in Blair and Kudrolli experiments \cite{blair}. When the bead number increases, the barrier seems to vanish.  Since the energy of an infinite chain is expected to coincide with the energy of the infinite ring, the energy per particle decreases towards an asymptotic value $U_\infty/ N$ whatever the angle $\alpha$. This asymptotic value can be evaluated by 
\begin{equation}
{U_\infty \over N} = \lim_{N \rightarrow \infty} {U_0 \over N} = -2 \zeta(3) \approx -2.404
\end{equation}
where $\zeta$ is the Riemann zeta function. 

The major feature of the energy landscapes presented in Figure \ref{fig_chain}(b) is the presence of an energy barrier whose angular position $\alpha_{max}$  is marginally sensitive to $N$. One would expect that the ring will form when the interaction between both extremities of the chain reaches high values, i.e. when they are close to each other. This argument is in favor of a typical distance of interaction and therefore to an increasing angle $\alpha_{max}$ with $N$. The fact that a specific angle still exists for large $N$ values underlines that some global properties of the chain emerges, and that subtle long-range phenomena have to be taken into account. It should be also noticed that the bending process from a chain to a ring illustrated in Figure \ref{fig_chain}(a) changes drastically the topology of magnetic field lines at a large scale, i.e. at a scale larger than the magnetized spheres. Two pictures of field lines obtained in our simulations are shown in Figure \ref{fig_chain}(c) and (d), presenting respectively dipole and multipole topologies. In the former case, the long field lines emerge from one chain extremity and sink on the other one whatever the number of spheres. In the ring case, each sphere is the source and sink of field lines. The field line structure evolves therefore from a dipole ($\alpha=0^\circ$) to a multipole ($\alpha=360^\circ$) topology. 

\begin{figure}[ht]
\begin{center}
\includegraphics[width=7.0cm]{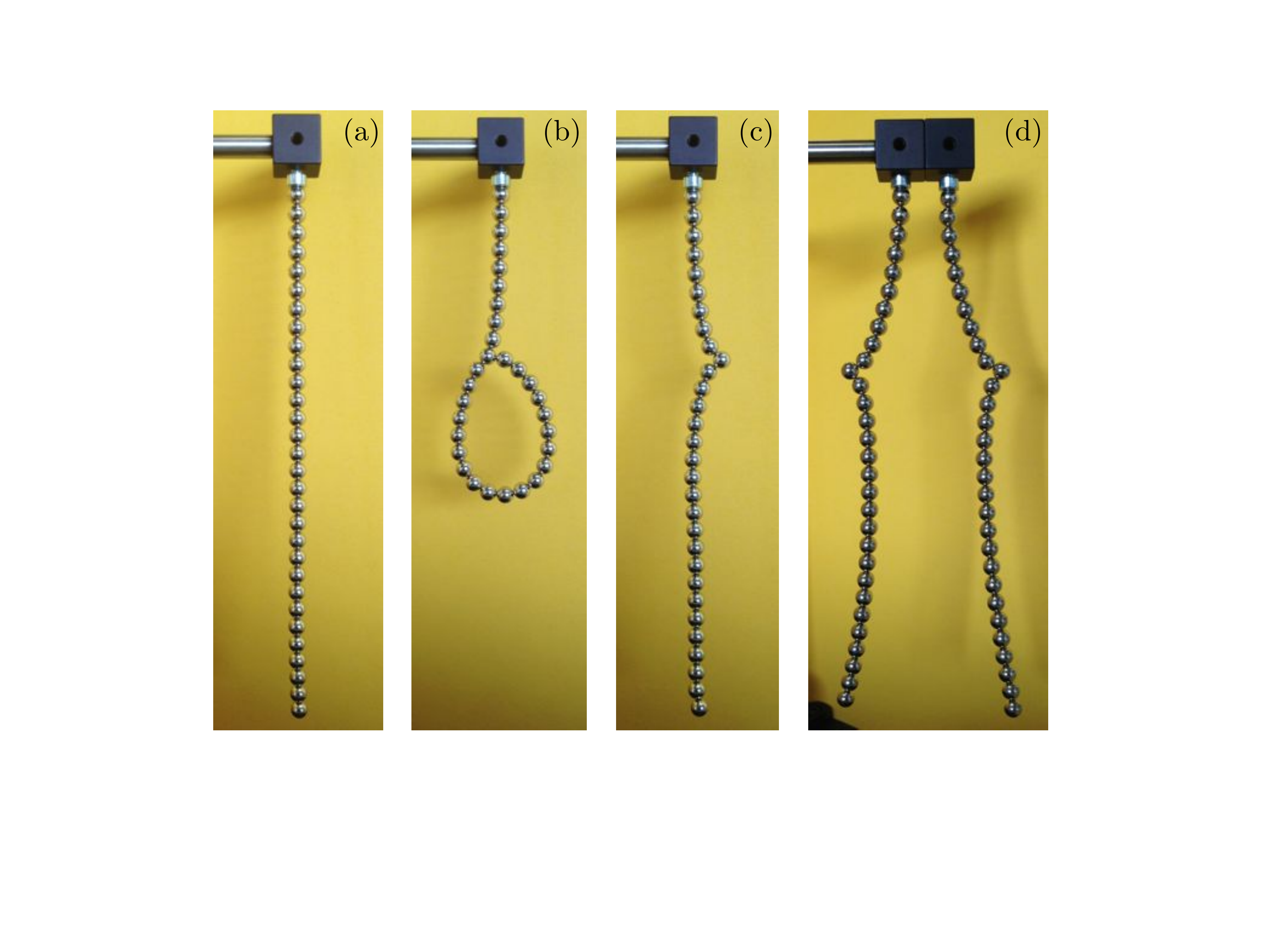}
\vskip 0.1cm
\includegraphics[width=8.0cm]{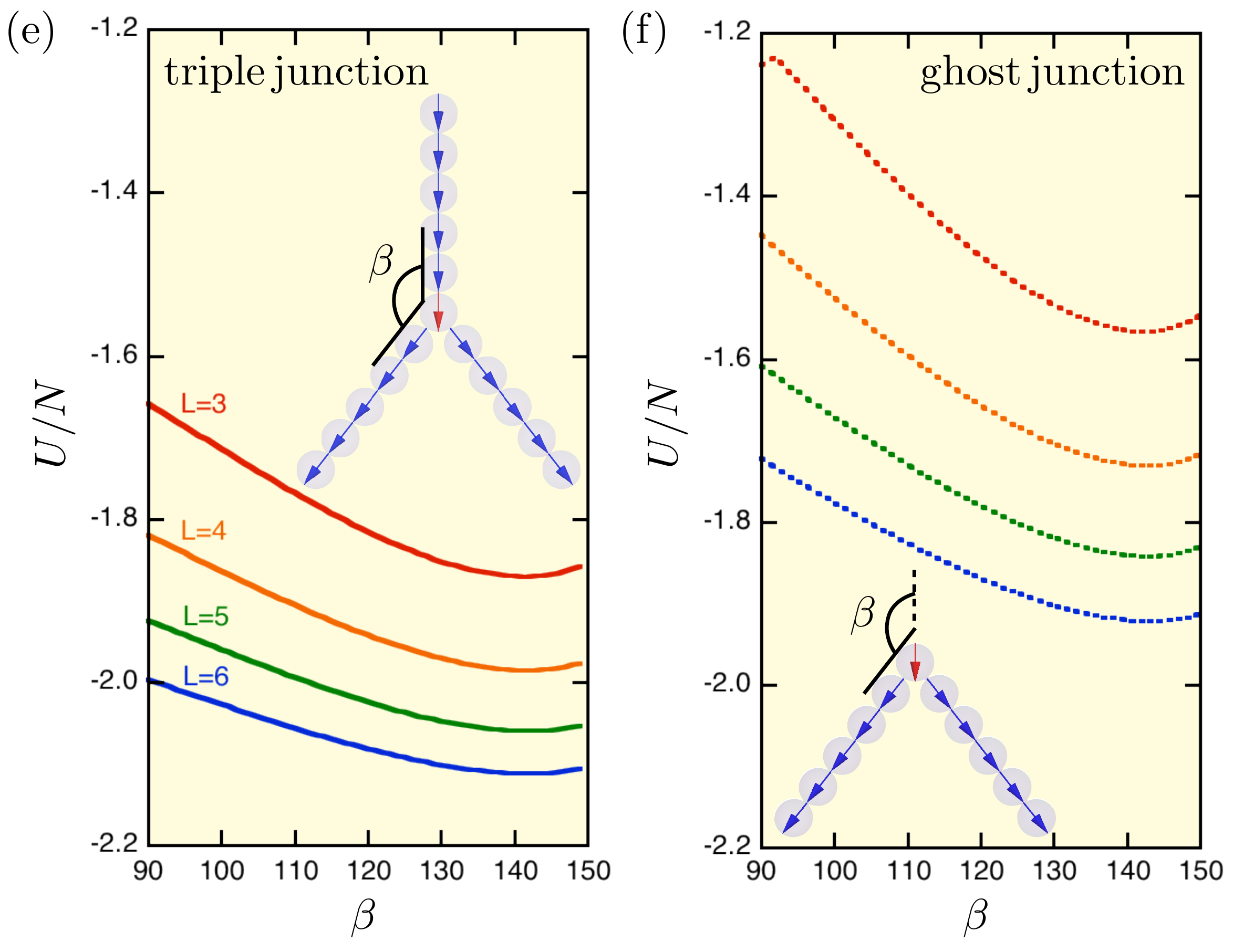}
\vskip -0.2 cm
\caption{ (a) Picture of a vertical chain, made of 30 magnetized spheres, attached at the top and submitted to gravity. (b) Same system forming a ``droplet" due to a magnetic reconnection. (c) Detaching a branch from the triple junction creates a ghost junction for which the memory of the dipole orientation is conserved. (d) Two ghost functions repel each other illustrating the fact that they possess identical magnetic charges.  (e) Potential energy $U/N$ as a function of the angle $\beta$ for a triple junction. Different branch lengths $L$ are illustrated ($N=3L+1$). A minimum is obtained for $\beta \approx 143^\circ$. The inset shows a triple junction with non equal angles minimizing the potential energy from numerical simulations. The red dipole corresponds to the junction itself and the branch length is $L=5$.  (f) Potential energy $U/N$ as a function of the angle $\beta$ for a ghost junction. Different branch lengths $L$ are illustrated ($N=2L+1$). The inset presents the ghost junction minimizing the energy as obtained from numerical simulations.  }
\label{fig_triple}
\end{center}
\end{figure}

For small $\alpha$ values and long chains, it can be shown that the leading terms of the potential are $U \approx U_0 + \alpha^2 / 4 N + ...$ such that any deviation from the rectilinear chain potential is proportional to $\alpha^2$ (see supplementary materials). Although the physical origin of this behaviour comes from the dipole-dipole interactions, this quadratic behaviour is shared by elastic systems. For a large number of beads, the chain is more flexible and a small energy input is able to bend the chain to overcome the barrier observed in Figure \ref{fig_chain}. Long chains are therefore forming rings or even ``droplets", as illustrated in Figure \ref{fig_triple}(b). A droplet is generated when one of the chain extremities touches a bead already connected to two neighbors. The reconnection of a chain into a droplet creates a junction where three branches meet. Such a triple junction is the focus of the present paper. 

One observes that the angles formed by the branches at the junction are non equal. Numerical simulations and experiments show that two angles are identical and larger than $120^\circ$, the third one being much smaller. Numerical simulations were performed to estimate those particular angles. Different triple junctions are considered. They are composed of a central bead with three branches each containing  $L$ beads for a total bead number $N=3L+1$. A variable angle $\beta$ is considered between a pair of branches. The energy landscapes are shown in Figure \ref{fig_triple}(e) as a function of $\beta$ between $90^\circ$ and $150^\circ$. Below $90^\circ$, unstable configurations are met, while above $150^\circ$, beads are overlapping. A minimum of $U/N$ is found for an angle $\beta \approx 143^\circ$, which is in agreement with our experimental observations. The inset of Figure \ref{fig_triple}(e) presents the dipole orientations for that configuration. The central dipole, illustrated in red, is seen to keep the orientation of the central branch.  

\begin{figure}[h]
\begin{center}
\includegraphics[width=8.5cm]{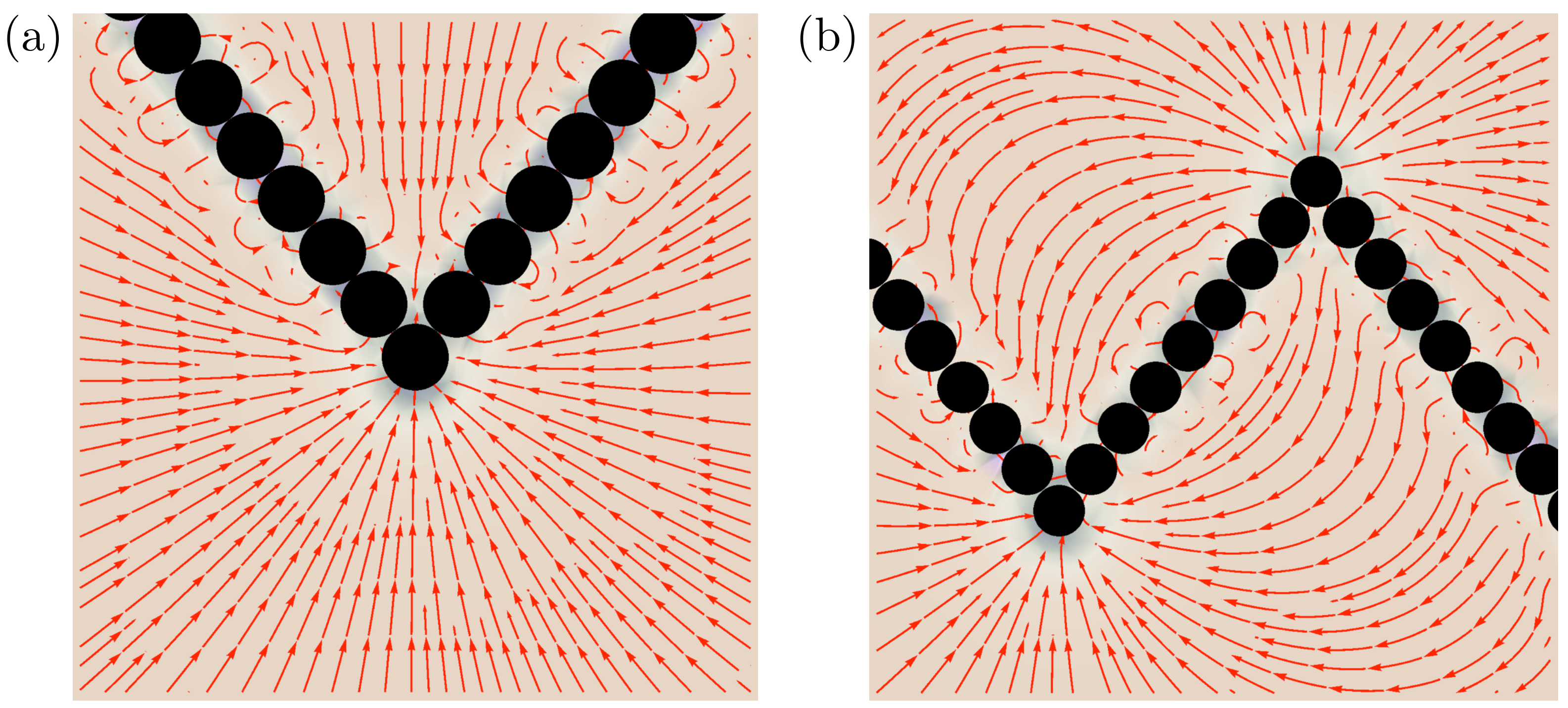}
\includegraphics[width=8.5cm]{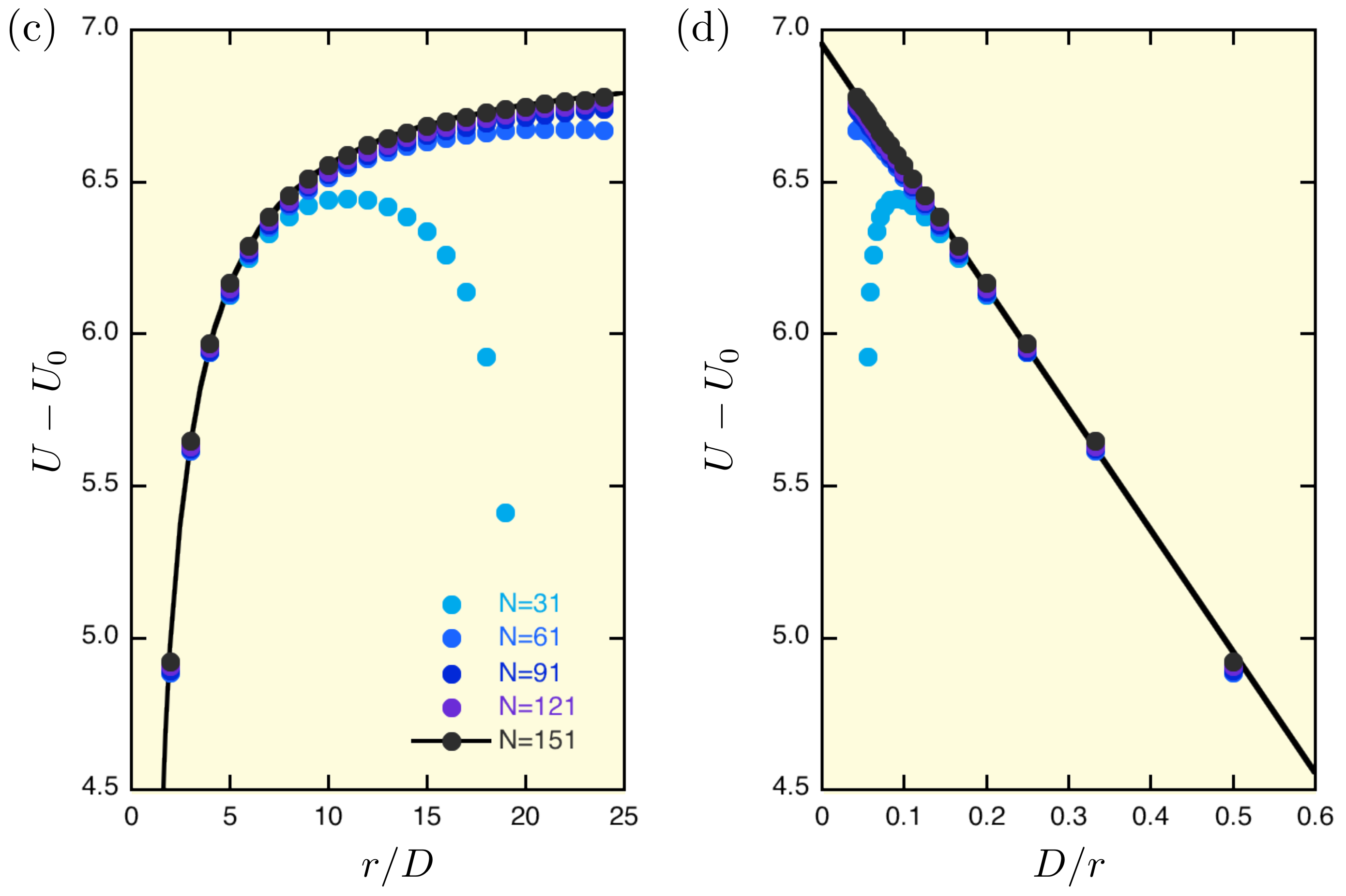}
\vskip -0.2 cm
\caption{(a) Field lines, as obtained from numerical simulations, around a ``magnetic ghost". The frustrated dipole behaves like a sink similarily to a monopole. (b) Two defects along a chain which are separated by $r/D=8$ bead diameters. Field lines seem to emerge from ghost junctions. (c) Dimensionless dipolar energy shifted by the chain energy $U_0$ for a 1D system containing two ``ghosts" as a function of the distance $r$ between them. Different system sizes are illustrated : $N=\{31, 61, 91, 121, 151\}$. The magnetic Coulomb interaction potential is fitted for the largest system size and is in excellent agreement with the data. (d) Same data as a function of $D/r$. In this plot, the Coulomb interaction is linear with a unique slope fitted on the data for $N=151$, providing ``magnetic charge" characteristics and the intercept provides twice the self-energy of a single ghost junction.
}
\label{fig_defect}
\end{center}
\end{figure}

A triple junction, resulting from a magnetic reconnection, is stable for non-equal angles $\{ 143^\circ,143^\circ,74^\circ \}$. Two branches are forming a pair against the third one. By removing the central branch, one expects that the pair of branches will reduce to a simple linear chain. The surprise is that the structure remains in the previous configuration with a similar angle $\beta$, keeping the memory of the triple junction !  This ``ghost junction" or ``magnetic ghost" is shown in Figure \ref{fig_triple}(c) while a sketch is given in Figure \ref{fig_triple}(f). The different steps for creating a magnetic ghost are given in the supplementary movie. One should note that gravity is not able to break the ghost junction in Figure \ref{fig_triple}, proving the mechanical stability of the newly formed structure. Figure \ref{fig_triple}(f) presents the energy per particle for that kind of configuration. A minimum is found for an angle around $\beta \approx 143^\circ$, close to the previous value. The remarkable feature of a ghost junction is that it should be associated to a chain in which dipole orientations suddenly change. For the bead which is the central point of the ghost junction, the dipole keeps the orientation of the branch which has been removed. This frustration should be attributed to a kind of defect between two domains of aligned dipoles. Figure \ref{fig_defect}(a) presents the field lines around a ``ghost", as obtained in numerical simulations. At the scale smaller than the bead diameter, the dipole nature of the components is observed near the chain. However, at a scale larger than the sphere diameter, the field lines converges towards the frustrated dipole. The latter seems to play the role of a monopole. This will be investigated below. 

Multiple ghost junctions can be created along a single chain. Zigzags are stable against gravity. If one takes a look at the magnetic field lines at a large scale around the zigzag, as shown in Figure \ref{fig_defect}(b), one discovers that the dipole organization along the zigzag is creating sources and sinks at the defects. Along the chain, the magnetic charges associated to successive ghost junctions have different signs due to the different orientation of the dipoles. Moreover, ghost junctions exhibit long range interactions. Figure \ref{fig_triple}(d) presents two ghost junctions on two different systems being attached to a support. Gravity orients the systems along the vertical direction. The systems are placed face-to-face for testing the interaction between ghost junctions. Due to their similar magnetic charges, the defects repel each other whatever the initial orientation of the systems. For different dipole orientations, ghosts attract each other, leading to a collapse of the system, not shown in Figure \ref{fig_triple}. 

The observation of both attractive and repulsive interactions between ghost junctions motivates a deeper analysis of such systems. Two pseudoparticles $A$ and $B$, acting as monopoles, are expected to be characterized by the Coulomb-like interaction potential 
\begin{equation}
u_{AB} = {\mu_0 \over 4 \pi} {Q_A Q_B \over r_{AB}} ,
\label{eq_coulomb}
\end{equation}
where the magnetic charge $Q$ is given by $Q=\pm m/\xi$ \cite{spinice}. The length $\xi$ provides a characteristic size for the pseudoparticle. This behaviour has been reported for spin ice systems \cite{spinice} which are geometrically frustrated ferromagnets on tetrahedral lattices. Spin ice systems lead to a fractionalization of dipoles into monopoles \cite{spinice}.

We studied the energy potential when the dimensionless distance $r/D$ between two ghost junctions is modified. The Figure \ref{fig_defect}(c) shows that dimensionless energy $U$ of the chain as a function of $r/D$. This energy is shifted by $U_0$ given by Eq.(\ref{eq_u0}), being the energy of a single chain containing the same number of beads. Different chain sizes are illustrated. For large systems, the interaction between successive defects is attractive and scales as $1/r$ : the interaction is remarkably Coulomb-like in between 2 and 25 sphere diameters. The agreement between the fit and the data is excellent. For small systems, finite size effects appear when $r/D$ has the same order of magnitude than $N$. This generates deviations from the Coulomb law. The same data are shown in Figure  \ref{fig_defect}(d) for different chain sizes and as a  function of $D/r$ for emphasizing the robustness of the Coulomb-like behaviour. For large $N$ values, all data  collapse on the same linear behaviour, meaning that $U-U_0$ measures the energy of the two interacting monopoles. The intercept with the vertical axis gives the self-energy of two isolated ghost junctions. Simulations give a dimensionless self-energy is $U-U_0 \approx 3.45$, close to what is expected from calculations (see supplementary materials). 

From the fit of the data of Figure \ref{fig_defect}(c-d) with equation (\ref{eq_coulomb}), taking into account the right units gives $\xi \approx 0.46 D$. This particular length should be attributed to the specific angle of the ghost junction and the associated dipole orientations.  One should remind that the system is composed of dipoles. It is therefore natural to obtain a characteristic length $\xi$ linked to the dipole size $D$, above which the system could be regarded as a macroscopic entity. Prior to the present work, spin ice systems were known to show monopoles at the microscopic level. Our observation of stable frustration at the macroscopic scale and for non-equilibrium systems opens ways to explore useful signatures of complex physical phenomena and in particular fractionalization.

We observed that monopoles emerge from a defect in the magnetic structure, exactly like in spin ice systems \cite{spinice}. From a general point of view, it is interesting to note that the boundary between two ordered domains has a spatial length that is smaller than the elementary magnetic cell, i.e. the size of bead in this case, the distance between two atoms in spin ice. In consequence, the origin of the mechanism for the obtention of monopoles resides in the fractionalization of dipoles at a scale close to the lattice unit, being the bead size here.

In summary, the organization of athermal magnetized spheres leads to a wide variety of structures. We have explored 1D structures such as chains, rings and junctions. By following simple processes such as bending, unexpected mechanically stable structures were discovered. We call them ``ghost junctions" because they keep the memory of a part of the system which has been removed. They seem to act as magnetic monopoles, as demonstrated by field lines and a Coulomb-like interaction. We hope our work will encourage experiments in magnetic dissipative systems like magnetic colloids \cite{fermigier,maret} and magnetic granular systems \cite{blair,tunable}.


\begin{acknowledgments} 
This work is financially supported by the University of Li\`ege (Grant FSRC-11/36). SD thanks FNRS for financial support. B.Vanderheyden and J.Martin are acknowledged for fruitful discussions. 
\end{acknowledgments} 




\begin{thebibliography}{99}

\bibitem{selfassembly} G.M. Whitesides and B. Grzybowski, Science {\bf 295}, 2418-2421 (2002)

\bibitem{pelesko}J.A.Pelesko, {\it Self-Assembly}, (Chapmann \& Hall, Boca Raton, 2007)

\bibitem{pnas} M.Boncheva, S.A.Andreev, L.Mahadevan, A.Winkleman, D.R.Reichman, M.G.Prentiss, S.Whitesides, and G.M.Whitesides, PNAS {\bf 102}, 3924-3929 (2005)

\bibitem{softmatter} G.Lumay, N.Obara, F.Weyer, N.Vandewalle, Soft Matter {\bf 9}, 2420 (2013)

\bibitem{aranson} A.Snezhko and I.S.Aranson, Nature Mat.  {\bf 10}, 698 (2011)

\bibitem{fermigier} R.Dreyfus, J.Baudry, M.L.Roper, M.Fermigier, H.A.Stone and J.Bibette, Nature {\bf 436}, 862 (2005)

\bibitem{maret} K. Zahn and G. Maret, Phys. Rev. Lett. {\bf 85}, 3656 (2000)

\bibitem{kadanoff} L.P.Kadanoff, {\it Statistical Physics : statics, dynamics and renormalization} (World Scientific, 2000) 

\bibitem{blair} D.L. Blair and A. Kudrolli, Phys. Rev. E {\bf 67}, 021302 (2003)

\bibitem{tunable} G. Lumay and N. Vandewalle, New J. Phys. {\bf 9}, 406 (2007)

\bibitem{carvente} O.Carvente, G.G.Peraza-Mues, J.M.Salazar, J.C.Ruiz-Suárez, Granular Matter {\bf 14}, 303-308 (2012).

\bibitem{falcon} E.Falcon, J.C.Bacri and C.Laroche, preprint arXiv 1306.4488 (2013).

\bibitem{jackson} J.D.Jackson, {\it Classical Electrodynamics}, 3rd ed (Wiley, New York, 1998)   

\bibitem{rosenweig}  R.E. Rosensweig, {\it Ferrohydrodynamics} (Dover Publications, 1997)

\bibitem{spinice} C.Castelnovo, R.Moessner and S.L.Sondhi, Nature {\bf 451}, 42 (2008)

\end{thebibliography}
\end{document}